# SCALABLE VOLTAGE-STABILITY DATASET GENERATION VIA BOUNDARY-PROXIMITY INDICATORS CLUSTERING

***Rock Agon*[*], Robin Preece, Jovica V. Milanović***

*Department of Electrical and Electronic Engineering, The University of Manchester, Manchester, UK*
**rock.agon@postgrad.manchester.ac.uk*

**Abstract**

This paper proposes a scalable framework for voltage-stability dataset generation. Voltage-stability-constrained planning increasingly relies on machine-learning surrogates but training them requires large datasets labelled by continuation power flow (CPF) results, which is computationally costly. To address this, this paper proposes a framework that uses hierarchical clustering on boundary-proximity indicators to reduce the number of required CPF evaluations. The proposed approach combines (i) uniform sampling of feasible operating space using Hit-and-Run Markov Chain Monte Carlo, (ii) structured stress directions via maximin Latin hypercube sampling (LHS), (iii) sensitivity-guided perturbations to target weak buses, and (iv) clustering-based representative CPF labelling that reconstructs the voltage stability margins of unlabelled operating points from representative cluster medoids. Results on the IEEE 39-bus system show that the proposed framework significantly reduces CPF evaluations by 95.45% while preserving high accuracy and boundary fidelity for both regression and classification tasks. The reduced surrogates remain structurally consistent with their full-CPF dataset counterparts, demonstrating the suitability and scalability of the proposed approach for operation and planning optimization.

## 1 Introduction

Voltage stability is a fundamental requirement in power system operation. It ensures that the system maintains acceptable voltage levels under normal and disturbed conditions. In modern grids, increased loading and the growing penetration of inverter-based resources significantly alter reactive power support and system dynamics, thereby exacerbating voltage stability challenges [1]. Theoretically, voltage instability originates from the nonlinear nature of power flow equations and is commonly associated with saddle-node bifurcations, where the equilibrium ceases to exist [2]. In practice, this leads to progressive voltage decline driven by reactive power deficiencies and adverse load behaviour. Consequently, accurate assessment of voltage stability margins (VSM) is essential for both operation and planning and performed through CPF. It traces an equilibrium trajectory beyond singularities and identifies the VSM [3].

Despite its accuracy, CPF is computationally intensive, particularly when applied across large sets of scenarios. This limitation becomes more critical in data-driven approaches, where machine learning (ML) models require large, labelled datasets to learn the mapping between system states and voltage security outcomes [4]. Among various ML, such as decision trees (DTs), support vector machines, neural networks, and ensembles, DT-based approaches are particularly attractive due to their interpretability and ability to extract explicit rules [5], [6]. However, their effectiveness is fundamentally tied to the quality of the training data [7].

Existing dataset generation methods often rely on random or guided sampling strategies, including gapsplit, importance sampling, and directed walks to improve the coverage of the near-boundary regions [4], [8]. More recent approaches incorporate sequential or probabilistic sampling to better explore the feasible space under uncertainty [7], [9]. Nevertheless, these methods still depend heavily on CPF-based labelling, making the dataset generation process computationally costly.

This paper leverages clustering techniques to group operating points that respond similarly to stress, using three indicators: the lowest bus voltage, the smallest reactive-power reserve, and the highest bus voltage sensitivity to reactive disturbances. Within each cluster, CPF is run on only one representative point under the N–1 contingency, and its voltage stability margin is assigned for every other point in the cluster. Specifically, the main contributions of this paper are:

- A framework for voltage-stability dataset generation, combining uniform sampling with precise sensitivity-based stress toward the stability boundary
- A clustering-based labelling strategy that runs CPF on a few representative points, reducing CPF evaluations without significantly losing voltage boundary fidelity.
- Application of a surrogate-to-polyhedra conversion to produce constraints as explicit linear inequalities..

## 2. Methodology

### *2.1 Proposed voltage-stability dataset generation technique*

The dataset generation approach comprises three stages: (i) sampling of feasible operating points across the dispatch space, (ii) sensitivity-guided stress augmentation to enrich the dataset, and (iii) voltage-stability labelling under a critical N–1 contingency. Each operating point is represented by a D-dimensional feature vector capturing the full pre-contingency network state, such as active and reactive power output of all generators, nodal active and reactive power demand, voltage amplitude and angle, and active and reactive power flow of all lines, together with the corresponding post-contingency VSM.

### 2.1.1 Operating Points Sampling through Hit-and-Run

The goal of this stage is to produce a large set of operating points of the power system, spread across plausible dispatch and loading conditions, so that the ML model sees the full diversity of states the grid can occupy.

The sampling used in this work follows the method in [7] to obtain representative operating states.

The operating points are sampled from the DC-feasible region of the system, defined by generator limits, load bounds, the power-balance equality, and transmission line limits. Together, these conditions define a convex polytope of feasible operating points that admits uniform sampling, which an AC-feasible set would not.

However, drawing uniform samples from this polytope directly is inefficient because of the geometry of the power balance equality in the polytope and the elongated nature of the polytope. Therefore, the power-balance equality is absorbed into the coordinate system through a singular value decomposition, so that every sample satisfies it by construction; and the remaining region is reshaped into a roughly ball-shaped polytope using the John-ellipsoid algorithm, so that a random walk can explore it efficiently in every direction.

Uniform samples are then drawn from the reshaped polytope using a random walk known as Hit-and-Run. Instead of choosing a fully random direction at each step, the walk is restricted to coordinate axes, which is computationally cheaper and equally effective once the region has been reshaped.

Each sample produced so far is only DC-feasible. It is therefore converted into a complete AC operating state by an ACOPF that finds the closest AC-feasible point to the sampled DC dispatch. Samples that fail AC convergence are discarded; the rest are kept as seeds for the next stage.

### 2.1.2 Stress Augmentation via Sensitivity-Guided Sampling

To improve coverage of the boundary, each seed has to be stressed along sensitivity-guided loading directions [8]. Therefore, for every seed, Nc candidate load stress directions are drawn from a maximin LHS. Nc has to be large enough to cover the load-direction space with the angular separation required by the maximin criterion, yet small enough to keep the per-seed sensitivity scoring step inexpensive. Each direction, which represents a vector of weights on the load buses, is scored by perturbing the seed load by:

$$\Delta Pd = \delta \, |dk| \odot Pd(s) \quad (1)$$

with $\delta$ the perturbation factor). Then, the power flow is resolved, and the voltage drop per unit perturbation is measured. The M highest-scoring directions are retained per seed. M has to be large enough to capture the dominant stress modes typically present in the system without inflating the computation cost, since every retained direction triggers an ACOPF-CPF pipeline during labelling. Then, the system load is stressed by the amplitude $\beta$ along those directions dm:

$$Pd(s,m) = Pd(s) + \beta \, |dm| \odot Pd(s) \quad (2)$$

$\beta$ is sampled differently depending on the decision tree:

*Classification tree:* a classifier separates operating points by the boundary value, so the most informative region is the High Information Content zone surrounding the boundary; the sampling therefore targets this zone, placing at least 75% of the points within it [8]. So $\beta$ can be defined as:

$$\beta \approx 0.8 + 1.5 \, Beta(5,2) \quad (3)$$

*Regression tree:* The goal is to learn the functional mapping from the operating state to the VSM. Hence the dataset must cover the entire feasible range of margins uniformly, not only the boundary region. To cover the feasible range of margins uniformly, the maximum amplitude $\beta_{max}(s,m)$ that drives the system to collapse is first found, then β is sampled uniformly:

$$\beta \approx U(0, \beta max) \quad (4)$$

### 2.1.3 Proposed Labelling Technique: Representative Medoid CPF

Performing CPF for every operating point is computationally prohibitive and the dominant cost in the entire pipeline. This section introduces the central contribution of the paper: a clustering-based labelling strategy that reduces the number of CPF evaluations by an order of magnitude while preserving voltage stability boundary fidelity. The idea is to group operating points that respond similarly to stress, run CPF only on one representative point per group, and propagate its margin to the rest. Specifically, hierarchical clustering is used to approximate the VSM of an operating point based on its pre-contingency power flow output.

After each AC optimal power flow, three boundary-indicators based on classical voltage stability theory are recorded:

- $Vmin$ = lowest voltage level in the system,
- $minQres$ = lowest reactive power reserve reflecting reactive power scarcity in the system,
- $maxVQsens$ = highest bus voltage sensitivity to Q changes, obtained from the reduced Jacobian matrix.

They form a compact vector z per operating point:

$$z = [Vmin, minQres, maxVQsens] \quad (5)$$

The operating points mapped into the standardized feature space z are grouped into clusters using hierarchical clustering with Ward linkage. After the clustering, each cluster Ck is represented by its medoid $m_k$, defined as:

$$mk = argmin \sum_{i,j \in Ck}^{n} \|zi - zj\|_2 \quad (6)$$

In data-driven voltage security assessment, the ML models are trained to recognize and prevent voltage instability or blackouts if any single critical component unexpectedly fails. The contingency is identified as the most critical for voltage stability by ranking outages according to their VSM.
For each medoid, the contingency is applied, and CPF is executed to compute the post-contingency VSM. CPF is performed with adaptive step sizes and a loading target equal to typically twice the base-case loading level.
In that way, a full dataset is created using only few CPF.

### *2.2 Existing voltage-stability dataset generation technique*

Conventionally, after every optimal power flow (Section 2.1.2), operating point is labelled independently: the critical N–1 contingency is applied and CPF computes the post-contingency VSM. The strategy is accurate but its cost scales linearly with dataset size, with CPF dominating the total. It serves as the comparative baseline for the proposed labelling.

### *2.3 Surrogate learning the voltage-stability rules*

Decision trees are trained on the datasets to learn an interpretable mapping between operating conditions and voltage stability margins. Two separate models are constructed based on their dataset: a regressor predicting the VSM value itself and a classifier predicting whether the VSM is higher than a predefined threshold. Both models are sparse oblique to capture the tilted geometry of the voltage-stability boundary more faithfully than axis-aligned splits, while remaining interpretable and tractable for optimization purpose.

#### *2.3.1 Sparse Oblique Regression Tree (SORT)*

SORT partitions the operating-state space into polyhedral regions and assigns a constant margin estimate to each. Each internal node defines a split $w^{T}x \leq t$, where w has at most $n_{sparse}$ non-zero entries and t is a scalar threshold. The sparsity constraint keeps splits interpretable and preserves the structure of the downstream MILP embedding. At each node, the $n_{sparse}$ candidate features are ranked by absolute correlation with the stability margin. A Ridge regression of VSM on these features yields the oblique direction w, which is $\ell_2$-normalised. This direction approximates the local axis of steepest margin variation, providing a more informative split than any single feature. The threshold t is selected by scanning the sorted projections $z_i = w^{T}x_i$ for the value that most reduces the variance of the VSM across the two resulting partitions. The gain is then multiplied by a balance factor 1 + H(pL), where H is the binary entropy of the left-child proportion, which favours balanced splits and prevents isolating small high-margin clusters.

#### *2.3.2 Sparse Weighted Oblique Classification Tree (SWOCT)*

SWOCT replaces axis-aligned splits with sparse oblique hyperplanes $w^{T}x \leq t$, where $w \in \mathbb{R}^{p}$ has at most $n_{sparse}$ non-zero entries. Each split is obtained in three steps. First, the $n_{sparse}$ candidate features are selected by ranking on a convex combination of absolute correlation with the binary label, standard deviation, and a small random component that injects exploration. Next, a class-balanced regularized logistic regression on these features yields a separating direction w. Finally, the threshold t along $w^{T}x$ maximizes the gain in information weighted by entropy, the reduction of impurities gain scaled by 1 + H(pL) to penalize unbalanced splits. Each leaf is labelled by majority vote, and the probability of the leaf is retained for a confidence-based prediction interface.

#### *2.3.3 Extraction of polyhedral stability regions*

The learned trees are transformed into explicit mathematical rules suitable for optimization and interpretation. Each root-to-leaf path defines a sequence of linear inequalities based on the splits. For leaves classified as secure, these inequalities define a convex polyhedron:

$$Pi = \{ x \mid Aix \leq bi \} \quad (7)$$

Extraction constructs (Ai, bi) by traversing each path and converting inequalities into standard linear form. The overall safe region is the union of convex polyhedra.

## 3 Results

### *3.1. Implementation details*

The proposed framework is applied to the IEEE 39-bus New England system. Table 1 summarises the main decision variables, parameters and hyperparameters used throughout the pipeline, as described and justified in Section II.

TABLE 1: Implementation decisions.

| Variable | Value |
|---|---|
| Dataset features dimensions D | 230 |
| Load bounds | [0.7, 1.6] × base load |
| Candidate load stress directions Nc | 30 |
| Retained top directions per seed M | 4 |
| Load perturbation factor δ | 2% |
| Contingency considered | Line 6 – 7 |
| Tree depth d; Sparsity per split $n_{sparse}$ | 4 ; 4 |
| Classifier boundary (threshold) | 5% [10] |

### *3.2 Dataset size*

A dataset size is not fixed in advance. It is determined empirically by growing the dataset iteratively until further additions produce no measurable improvement in the surrogate quality. Two datasets were then iteratively generated: one for SORT and one for SWOCT. Three convergence criteria were applied jointly to decide on the dataset size. For the classifier, sampling continued until: (i) at least 75% of operating points fell within the boundary band, ensuring the decision boundary was well represented; (ii) the cross-validation accuracy stagnated below an improvement of 0.1% per additional 1,000 points; and (iii) accuracy exceeded 90% on a held-out test set. For the regressor, sampling continued until: (i) the generated points covered the full feasible range of VSMs; and (ii) the validation MAE stagnated below an improvement of 0.0001 per additional 1,000 points; and (iii) the $R^2$ exceeded 0.98 on a test set. With these criteria, both datasets (for classifier and for regressor) converged at approximately 22,000 points.

### 3.3 *Clustering quality*

This section evaluates whether the Ward hierarchical clustering successfully group operating points that share similar VSM and quantifies the labelling error introduced by propagating each medoid's margin to its cluster members. This evaluation is independent of the decision tree type.

*3.3.1 Correlations:* The correlation between distances in the indicator space and variations in the VSMs is first investigated. For each of the three features and for their combined Euclidean distance, pairwise feature distance is plotted against the corresponding normalized CPF-margin distance. All three features show a monotonic relationship: $minQres$ is the strongest predictor ($\rho = 0.81$), followed by $Vmin$ ($\rho = 0.70$), while $maxVQsens$ is the weakest ($\rho = 0.53$). The combined distance achieves $\rho = 0.78$, confirming that the points close in clustering feature space tend to share a similar margin.

*3.3.2 MAE-vs-K Analysis:* Since medoid VSMs are propagated to all points within a cluster, the labelling error is quantified before surrogate learning. Operating points are stratified by proximity to the stability boundary into near-boundary (40%), intermediate (30%), and far-from-boundary (30%) regions.

Fig. 1 reports the mean absolute error (MAE) between propagated and CPF-computed VSMs as the number of clusters K increases. The near-boundary MAE remains low across all resolutions, decreasing below 0.002 at K = 1000. This is particularly important because near-boundary operating points determine the binding constraints in voltage-stability-constrained planning. The intermediate MAE saturates around 0.004 beyond K = 800, close to the global MAE. The largest and instable errors occur in the deep-interior region (far-from-boundary MAE ≈ 0.008), but these points are typically non-binding in operation or planning optimization.

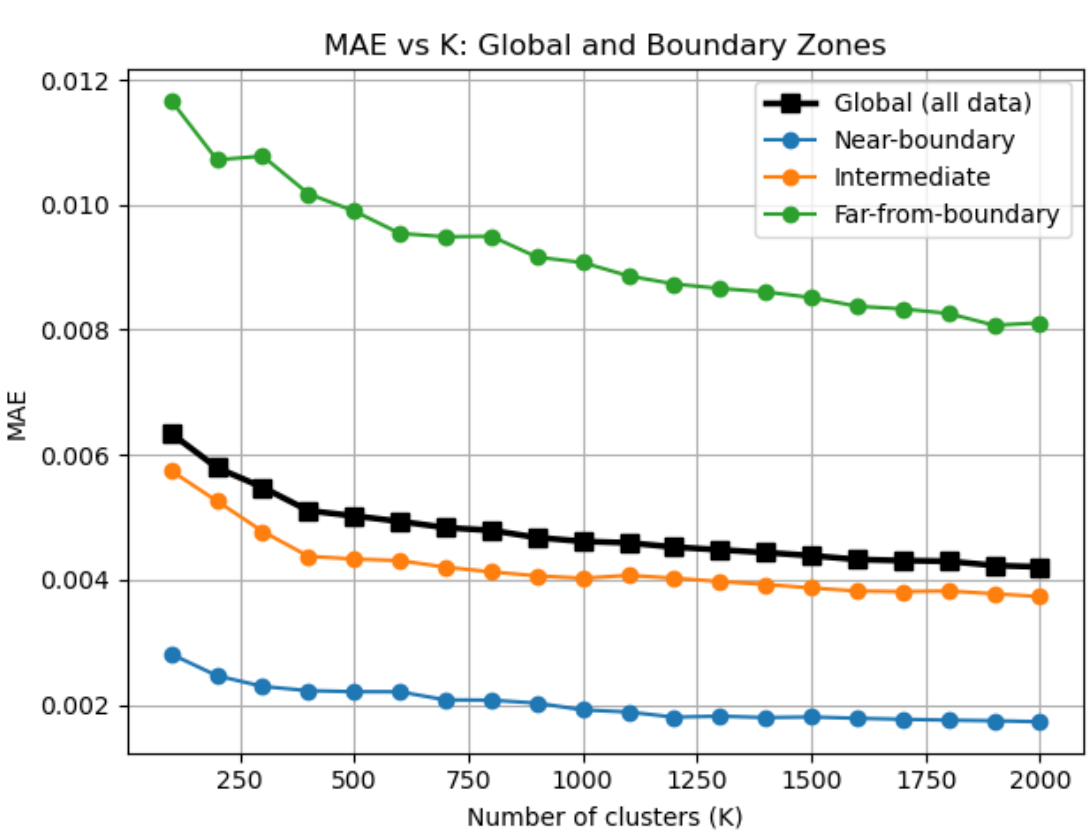

Fig. 1: Clustering errors for different boundary regions.

Although improvements beyond K ≈ 400 are marginal, K = 1000 was chosen to ensure sufficient resolution for both tasks, particularly for classification, where small VSM perturbations near the security threshold can flip the binary label. As shown in Fig. 2, the selected 1000 clusters preserve strong agreement between propagated and CPF-computed margins, with larger residuals appearing only at high VSM values.
Table 2 summarizes the two dataset's sizes and the computation cost reduction induced by the proposed approach.

TABLE 2: Comparison of full and reduced-CPF datasets.

| Property | Full | Reduced-CPF |
|---|---|---|
| Operating points (labelled) | 22,000 | 22,000 |
| CPF evaluations | 22,000 | 1,000 |
| CPF cost (relative) | 1× | 0.045× |
| Labelling strategy | per-point | medoid-propagated |

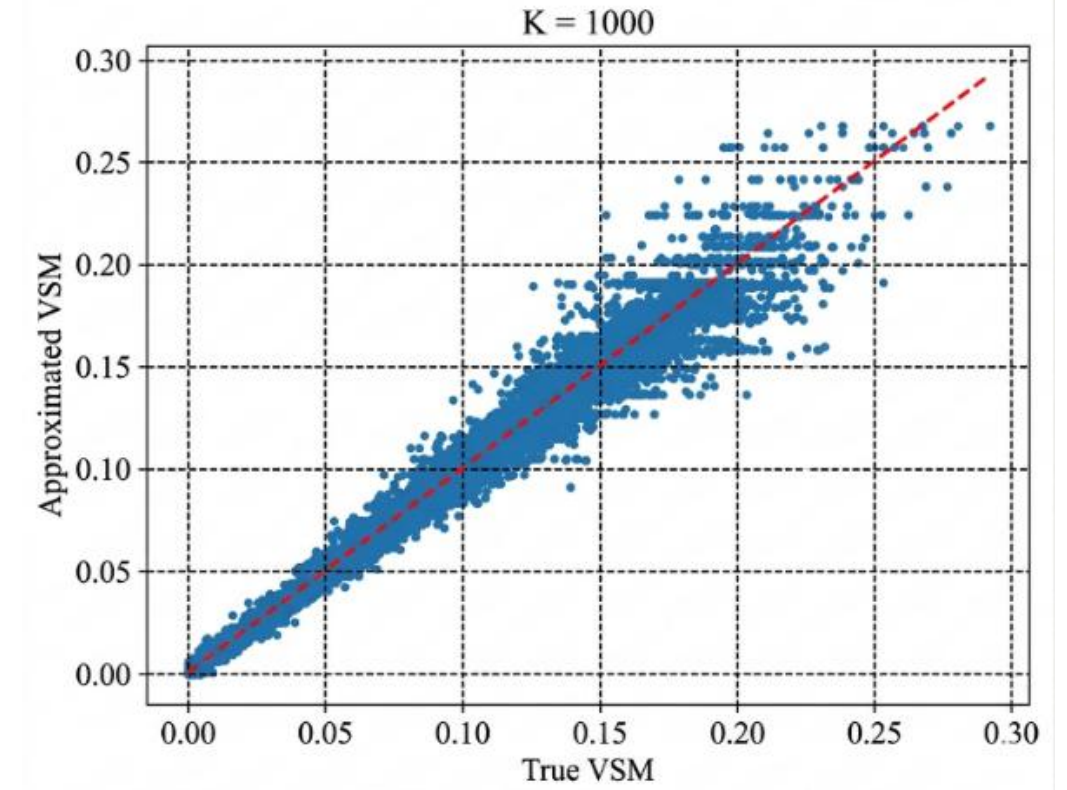

Fig. 2: Approximated versus CPF-computed VSMs.

### 3.4 *Surrogate modelling quality*

*3.4.1 Performances:* Four surrogate models were constructed:

- SORT using the full dataset,
- SORT using the reduced-CPF dataset,
- SWOCT using the full dataset,
- SWOCT using the reduced-CPF dataset.

Table 3 summarizes the hold-out test performance.

TABLE 3: Hold-out test-set performances

| Regression Models | MAE | RMSE | $R^2$ |
|---|---|---|---|
| Full Dataset | 0.0036 | 0.0053 | 0.9895 |
| Reduced-CPF Dataset | 0.0048 | 0.0071 | 0.9822 |
| Absolute error | 0.0012 | 0.0018 | 0.0073 |

| Classification Models | Accuracy | Precision | F1-score |
|---|---|---|---|
| Full Dataset | 0.9623 | 0.9618 | 0.9641 |
| Reduced-CPF Dataset | 0.9289 | 0.9677 | 0.9151 |
| Absolute error | 0.0475 | 0.0059 | 0.049 |

The full-dataset SORT achieves an MAE of 0.0036 with a $R^2$ score of 98.95, indicating that the sparse oblique partition successfully captures the nonlinear geometry of the voltage stability surface. Similarly, the full-dataset SWOCT achieves an accuracy of 96.23% with 96.18% precision, demonstrating strong separability between secure and insecure regions. More importantly, the reduced-CPF dataset surrogates maintain performance very close to their full-CPF counterparts despite requiring only 1,000 CPF evaluations during dataset generation. The reduced-CPF dataset SORT exhibits only a marginal degradation in MAE (0.0048 versus 0.0036), while the reduced-CPF dataset SWOCT preserves high precision (96.77%) and good classification accuracy (92.89%).

These results confirm that the proposed reduction framework preserves the dominant stability structure required for surrogate learning. The small performance degradation is consistent with the clustering approximation error and remains substantially smaller than the achieved computational savings.

*3.4.2 Boundary visualization:* Fig. 3 and 4 visualize the learned stability regions in a two-dimensional operating subspace defined by the reactive power outputs of generators 6 and 8. Although the surrogates operate in high dimensional feature space, these two variables are selected for visualization. The red areas represent insecure regions while the green the secure one. Any other couple of generators would show different repartitions of the VSM in their space, but the general trend would remain the same. Fig. 3 presents the regression-based stability surface. The predicted VSM varies smoothly across the operating space, with progressively smaller margins appearing as reactive support weakens. By contrast, Fig. 4 illustrates the binary secure/insecure regions produced by the classifier at the boundary chosen at VSM = 0.05 according to WECC in [10].

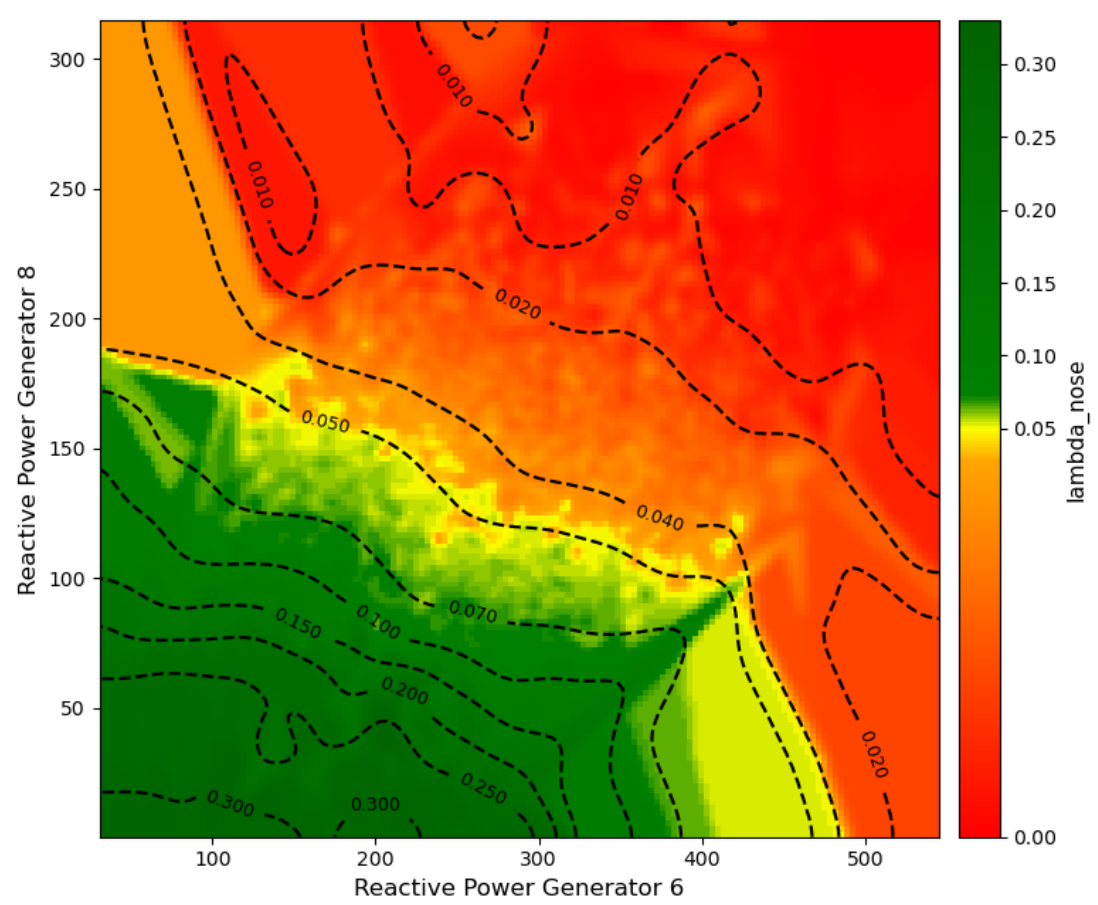

Fig. 3: Regression-based boundary.

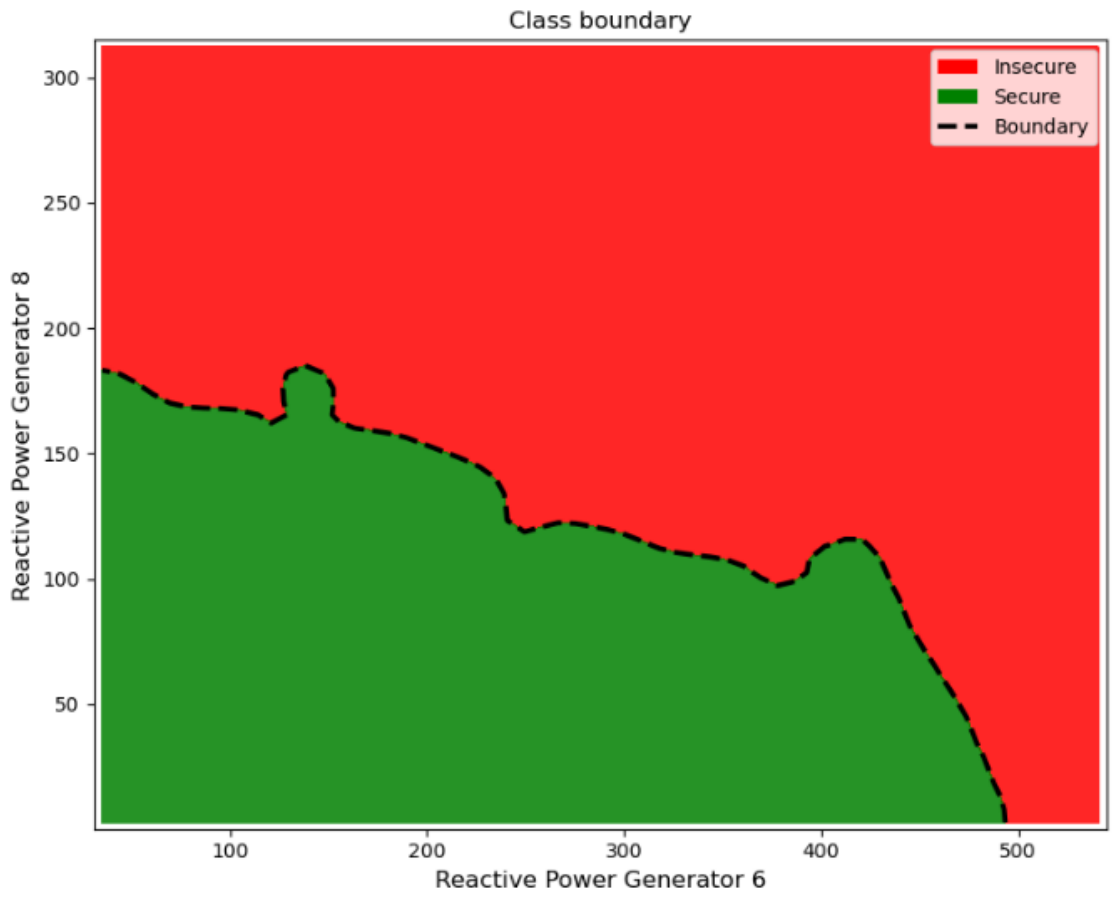

Fig. 4: Classification-based boundary.

*3.5 Full vs. reduced-CPF surrogates: preservation of the decision geometry*

Both reduced-CPF dataset surrogates match the predictive accuracy of their full dataset counterparts. In stability-constrained planning the surrogate is embedded through its root-to-leaf inequalities rather than queried point-by-point. The relevant question is therefore whether the two datasets yield the same decision voltage stability geometry.

*3.5.1 SORTs Alignment:*

*a) Leaf-level:* Fig. 5 shows the predicted VSM per leaf. At the output level, the two regressors assign near-identical margins to each leaf, with the largest disagreement, $1.2\times10^{-2}$ at leaf 15, falls in the highest-margin region deep-interior points far from collapse, which do not enter the binding constraint set.

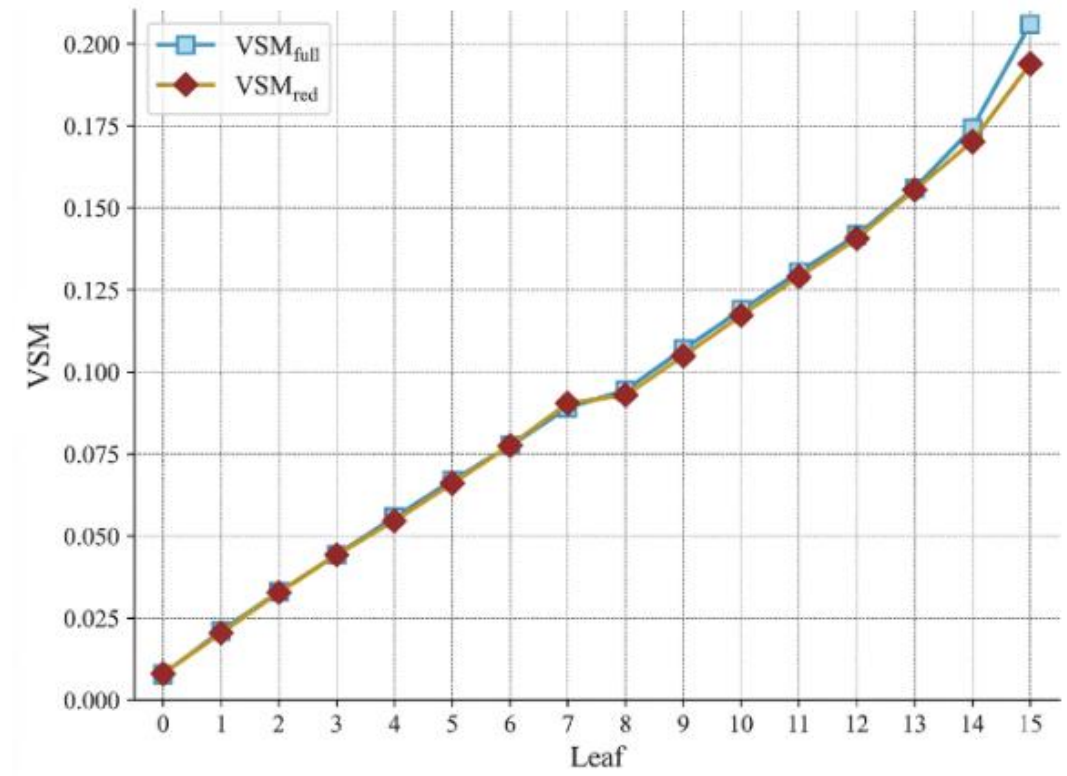

Fig. 5: Regressors leaf agreement.

*b) Splits:* At the mechanism level, the separators carving the boundary region share the same features with near-coincident geometry: the roots share all four features and this alignment holds through depth 2. In the root equation below, (F) represents the full-CPF tree and (R) the reduced-CPF tree.

$$F: 0.778V_{a6} - 0.129V_{a8} - 0.612V_{a11} - 0.055P_{g2} \leq 0.009$$
$$R: 0.780V_{a6} - 0.129V_{a8} - 0.610V_{a11} - 0.057P_{g2} \leq 0.010$$

Table 4 reports the per-depth alignment. Here cos(w_F, w_R) is the cosine of the angle between the two splits' normal vectors, measuring whether the full- and reduced-CPF trees orient their splits along the same direction in feature space (1 denotes identical orientation), while |t_F − t_R| is the difference in their thresholds, measuring any offset between the two parallel planes; together they quantify how closely the two splits coincide in orientation and position. Alignment then decays monotonically with depth, substituting variables within the same physical mode at depth 3 and diverging in variable family only at depth 4. Since shallow splits isolate the boundary and deep splits subdivide the secure interior, the divergences are confined to the non-binding region, consistent with the high-VSM regions in Fig. 5.

TABLE 4: SORTs split alignment (full vs. reduced-CPF).

| Depth | Shared | cos(wF, wR) | \|tF − tR\| |
|---|---|---|---|
| 1 (root) | 4/4 | 0.999 | 0.001 |
| 2 | 4/4 | 0.998 | 0.001 |
| 3 | 3/4 | 0.94 | 0.005 |
| 4 | 1/4 | 0.71 | 0.020 |

*3.5.2 SWOCTs Alignment:*

*a) Leaf-level:* As shown in Table 5, the two trees agree on 15/16 leaves. The only disagreement, leaf 3, lies on the decision boundary, where a small clustering error flips the label without altering the overall partition.

TABLE 5: SWOCTs leaf agreement.

| Region | Full | Red. | Region | Full | Red. |
|---|---|---|---|---|---|
| 0 | 0 | 0 | 8 | 0 | 0 |
| 1 | 0 | 0 | 9 | 0 | 0 |
| 2 | 0 | 0 | 10 | 0 | 0 |
| **3** | **1** | **0** | 11 | 1 | 1 |
| 4 | 0 | 0 | 12 | 0 | 0 |
| 5 | 0 | 0 | 13 | 1 | 1 |
| 6 | 0 | 0 | 14 | 1 | 1 |
| 7 | 1 | 1 | 15 | 1 | 1 |

*b) Splits:* At the mechanism level, both roots are driven by the same dominant feature, Qf,3→4. Qg,2 is substituted by Qf,10→11, both proxying the same reactive-corridor stress.

$F: 0.844Qf_{3\text{-}4} + 0.168Qg_3 - 0.418Qf_{10\text{-}32} + 0.29Qg_2 \leq -0.072$

$R: 0.838Qf_{3\text{-}4} + 0.157Qg_3 - 0.306Qf_{10\text{-}32} + 0.42Qf_{10\text{-}11} \leq -0.12$

Table 6 reports both exact feature overlap and physical-family agreement (voltage angles, magnitudes, or reactive corridors). At depth 2, the full-CPF tree reads the symptoms of reactive shortage (voltage sags at buses 3, 5, 6) while the reduced-CPF tree reads its causes (reactive-flow deficits on lines 6 → 11 and 16 → 19). At depth 3, both substitute angular variables within the same family (bus-9/39 vs. bus-22/35), while the right subtree stays near-identical. At depth 4 the trees diverge in variable family, the reduced-CPF tree introducing voltage magnitudes and reactive injections in place of angles.

TABLE 6: SWOCT split alignment (full vs. reduced-CPF).

| Depth | Exact | Same family | $\lvert tF - tR \rvert$ |
|---|---|---|---|
| 1 (root) | 3/4 | 4/4 | 0.05 |
| 2 | 1/4 | 4/4 | 0.02 |
| 3 | 2/4 | 4/4 | 0.01 |
| 4 | 0/4 | 2/4 | – |

### *F. Summary of Contributions*

The results confirm that the reduced framework is practically valuable for stability-constrained planning. The sampling-and-stress pipeline exposes the surrogates to the full geometry of the voltage-stability surface, which is the prerequisite for reliably deriving secure regions and embedding them into optimisation. By clustering operating points using boundary-proximity indicators, the framework substantially reduces CPF computations while preserving high fidelity near the stability boundary. Approximation errors are concentrated in deep-interior operating regions, which rarely define binding constraints in stability-constrained planning. The reduction is therefore boundary-faithful where it matters, and large enough to make the technique cheap to regenerate when topology or operating envelope changes.

A further observation concerns the two surrogate types. Regression surrogates tolerate propagated labelling errors well because VSM varies continuously across the operating space. Classifiers are more sensitive near the security threshold, where small VSM perturbations can flip the binary label. Nonetheless, the reduced-CPF classifier still preserves the dominant geometry of the secure region and is therefore usable for voltage stability constrained optimisations.

## 4. Conclusion

This paper proposed a scalable framework for voltage stability dataset generation using representative CPF approximation. By combining physics-informed sampling, hierarchical clustering, and medoid-based labelling, the proposed method significantly reduced CPF computations by 22x on the specific IEEE 39-bus system case, while preserving boundary fidelity. Results showed that the reduced-CPF datasets maintained high surrogate accuracy of 92% and strong structural consistency with their full-CPF dataset counterparts for both regression and classification tasks. The proposed framework, therefore, enables efficient and interpretable voltage stability assessment suitable for planning and operation optimizations. Future work will use the extracted regions into planning models.

## References

[1] N. Hatziargyriou et al., "Definition and classification of power system stability—Revisited and extended," IEEE Trans. Power Syst., vol. 36, no. 4, pp. 3271–3281, 2021.

[2] T. Cutsem and C. Vournas, Voltage Stability of Electric Power Systems. Boston, MA: Springer US, 1998.

[3] V. Ajjarapu and C. Christy, "The continuation power flow: A tool for steady state voltage stability analysis", IEEE Trans. Power Syst., vol. 7, no. 1, pp. 416–423, 1992.

[4] F. Thams, et al., "Efficient Database Generation for Data-Driven Security Assessment of Power Systems," IEEE Trans. Power Syst., vol. 35, no. 1, pp. 30-41, 2020.

[5] Q. Hou, et al., "Sparse Oblique Decision Tree for Power System Security Rules Extraction and Embedding," IEEE Trans. Power Syst., vol. 36, no. 2, pp. 1605-1615, 2021.

[6] C. Liu et al., "A Systematic Approach for Dynamic Security Assessment and the Corresponding Preventive Control Scheme Based on Decision Trees," IEEE Trans. Power Syst., vol. 29, no. 2, pp. 717-730, 2014.

[7] T. Xia, et al., "Database Generation for Data-Driven Power System Security Assessment Under Uncertainty," IEEE Trans. Power Syst., vol. 39, no. 5, pp. 6168-6182, 2024.

[8] Krishnan, et al., "Efficient Database Generation for Decision Tree Based Power System Security Assessment," IEEE Trans. Power Syst., vol. 26, no. 4, pp. 2319-2327, 2011.

[9] A.-A. B. Bugaje, et al., 'Split-based sequential sampling for realtime security assessment', International Journal of Electrical Power Energy Syst., vol. 146, p. 108790, 2023.

[10] WECC, 'TPL-001-WECC-CRT-4 Transmission System Planning Performance'.